\documentclass[twocolumn,aps,amssymb,showpacs,prb]{revtex4}
\usepackage{graphicx}
\usepackage{textcomp}

\begin{document}

\title{Magnetic-order induced phonon splitting in MnO from far-infrared spectroscopy}

\author{T.~Rudolf}
\author{Ch.~Kant}
\author{F.~Mayr}
\author{A. Loidl}
\affiliation{Experimental Physics V, Center for Electronic Correlations and
Magnetism, University of Augsburg, D-86135~Augsburg, Germany}

\date{\today}

\begin{abstract}
Detailed far-infrared spectra of the optical phonons are reported for
antiferromagnetic MnO. Eigenfrequencies, phonon damping and effective plasma
frequencies are studied as a function of temperature. Special attention is paid
to the phonon splitting at the antiferromagnetic phase transition. The results
are compared to recent experimental and theoretical studies of the spin-phonon
coupling in frustrated magnets, which are explained in terms of a spin-driven
Jahn-Teller effect, and to ab initio and model calculations, which predict
phonon splitting induced by magnetic order.
\end{abstract}



\maketitle

\section{Introduction}

Spin-phonon coupling in magnetic semiconductors has been an active area of
research for
decades.~\cite{baltensperger,brueesch,lockwood,wakamura,wesselinowa,rudolf72}
In correlated matter it has been revived and came into the focus of modern
solid state physics again, a fact that had been triggered by the observation of
the splitting of phonon modes just at the onset of antiferromagnetic (AFM)
order. This was investigated by far-infrared spectroscopy in geometrically
frustrated ZnCr$_2$O$_4$ (Ref.~\onlinecite{sushkov}) and CdCr$_2$O$_4$
(Ref.~\onlinecite{rudolf9}) as well as in bond frustrated ZnCr$_2$S$_4$
(Refs.~\onlinecite{rudolf9,hemberger97}) and ZnCr$_2$Se$_4$
(Refs.~\onlinecite{rudolf9,rudolf75}). In the latter compound AFM order can be
suppressed by moderate external magnetic fields~\cite{hemberger98} and the
phonon splitting fully disappears as the magnetic ordering temperature $T_N$
approaches 0~K (Ref.~\onlinecite{rudolf75}).

Of course by establishing long-range spin order at $T_N$, these chromium
spinels lower their symmetry and reveal slight distortions from cubic symmetry.
But these distortions are too small to explain the significant splitting of
phonon modes. In local spin-density approximations (LSDA) the non-cubic
behavior of the phonon properties can be explained in the absence of any
structural distortion and arises due to the anisotropy induced by the magnetic
order alone.~\cite{fennie} Furthermore, the structural distortions at the onset
of antiferromagnetic order in these and similar spinel compounds have been
explained by the concept of a spin-driven Jahn-Teller
effect.~\cite{yamashita,tchernyshyov}

The idea of a purely magnetic order induced phonon splitting has been put forth
by Massidda {\it et al.}~\cite{massidda} for the AFM transition metal
monoxides. The splitting of the transverse optical modes has been indeed
experimentally documented by Chung {\it et al.}~\cite{chung} in MnO and NiO by
inelastic neutron scattering. As the late transition monoxides are prototypical
correlated electron systems and benchmark materials for Mott-Hubbard
insulators~\cite{anisimov} it seems self-evident to reinvestigate MnO by
far-infrared spectroscopy. Despite the fact that a number of far-infrared
experiments on MnO have been published,~\cite{plendl,kinney,mochizuki}
temperature dependent measurements are absent and to our knowledge the
splitting of phonon modes below the AFM phase transition has not been reported
so far.

In principle it seems fascinating to compare the splitting of phonon modes in a
canonical antiferromagnet with strong exchange striction and only weak
frustration, with the splitting observed in strongly geometrically frustrated
magnets. MnO has a Curie-Weiss temperature of $\Theta_{CW} = -548$~K
(Ref.~\onlinecite{tyler}) and hence is characterized by a frustration parameter
$f = |\Theta_{CW}|/T_N \sim 5$. This value has to be compared with values in
strongly frustrated magnets, like the chromiums spinel ZnCr$_2$O$_4$, with
frustration parameters of the order of 20 (Ref.~\onlinecite{ramirez}).

At room temperature MnO is paramagnetic and exhibits the cubic NaCl structure
($Fm\bar{3}m$; $a = 0.44457(2)$~nm).~\cite{morosin} The transition into the AFM
state at $T_N = 118$~K (Ref.~\onlinecite{morosin}) is accompanied by a small
rhombohedral distortion,~\cite{tombs} with $a = 0.44316(3)$~nm and $\alpha =
90.624(8)^\circ$ at liquid helium temperature.~\cite{morosin} Despite the fact
that MnO basically is the simplest transition-metal monoxide with Mn$^{2+}$
having a half filled $d$-shell and hence a negligible spin-orbit coupling, at
the magnetic ordering temperature it exhibits an anomalous large
exchange-striction, which arises from the dependence of the exchange energies
on inter-ionic distances, and amounts $\Delta a/a \approx 1.1\times10^{-4}$
(Ref.~\onlinecite{morosin}). In the AFM phase the magnetic moments are arranged
in ferromagnetic (FM) sheets parallel to (111) planes which are coupled
antiferromagnetically.~\cite{shull} The moments lie within these FM
planes~\cite{shull,shaked} and are oriented parallel and antiparallel to the
$\langle11\bar{2}\rangle$ direction.~\cite{goodwin} It has been pointed out
that this spin structure is not compatible with the rhombohedral distortion.
The true lattice symmetry definitively should be lower.~\cite{shaked}

\section{Results and Discussion}

A high-quality single crystalline platelet of MnO with one side polished to
optical quality was purchased from MaTecK GmbH. The surface was parallel to a
(100) plane. The reflectivity measurements were carried out in the far- and
mid-infrared range using the Bruker Fourier-transform spectrometers IFS 113v
and IFS 66v/S which are equipped with a He bath cryostat. In the far-infrared
we measured the phonon spectra with high accuracy in a range from 50 to
700~cm$^{-1}$. Additional measurements up to 8000~cm$^{-1}$ were performed to
unambiguously determine $\epsilon_{\infty}$, which is governed by electronic
contributions only. The reflectivity spectra of MnO are reported as function of
temperature between liquid helium and room temperature.

\begin{figure}
\includegraphics{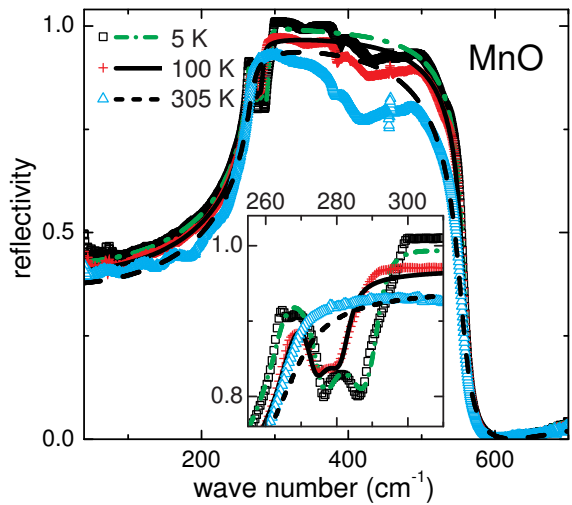}
\caption{\label{fig1}(Color online) Reflectivity spectra of MnO at 5~K, 100~K
and 305~K. The lines are results of fits based on a generalized oscillator
model. The inset shows the splitting of the phonon mode below $T_N$ into at
least three modes at 5~K. The fits at 5 and 100~K utilize three modes.}
\end{figure}

Representative results of the far-infrared reflectivity $R$ at 5~K, 100~K and
305~K are shown in Fig.~\ref{fig1}. At room temperature, as expected for
paramagnetic MnO, we find a broad reststrahlen spectrum due to transverse and
longitudinal optical phonons of the NaCl structure. At zero wave vector the two
transverse optical phonon modes are degenerate and determine the strong
increase of the reflectivity close to 250~cm$^{-1}$. The longitudinal optical
phonon is responsible for the steep decrease of the reflectivity at
550~cm$^{-1}$. In a harmonic solid, one would ideally expect that the
reflectivity in between these two characteristic frequencies, where $\epsilon'$
is negative, is close to 1. The observed structure in the reflectivity as
documented in Fig.~\ref{fig1}, especially the characteristic hump close to
500~cm$^{-1}$, probably results from two-phonon processes involving zone
boundary optical and acoustical modes, which sum up to a zero-wave vector
excitation with a dipole moment transferred from the transverse optical phonon
as discussed later in the text. Despite this fact and contrary to earlier work
we made no attempts to get a better fit involving two
eigenmodes,~\cite{plendl,mochizuki} but we will document that a reasonable fit
can be obtained with realistic values of one transverse and one longitudinal
optical mode.

As the temperature is lowered, the reflectivity in the frequency range of the
reststrahlen band slightly increases reaching values close to~1 at 5~K. At the
same time the hump-like feature becomes weaker while the overall shape of the
reflectivity spectrum roughly remains unchanged. Both facts indicate the
decreasing influence of anharmonicity when approaching low temperatures.
However, a closer inspection reveals that just at $T_N$ a clear dip develops in
the restrahlen band close to 280~cm$^{-1}$ which indicates the splitting of
modes. This mode splitting is shown in the inset of Fig.~\ref{fig1}, which
provides an enlarged view of the reflectivity between 255 and 310~cm$^{-1}$. At
100~K a dip evolves at 280~cm$^{-1}$ and at 5~K it seems that the structure of
the dip becomes even more complex indicating a splitting into three modes. This
will be analyzed in more detail later.

The reflectivity spectrum of MnO has been modeled with a generalized oscillator
model,~\cite{gervais} where the factorized form of the complex dielectric
constant, $\epsilon = \epsilon' + i \epsilon''$, is determined by transverse
and longitudinal eigenfrequencies and damping constants:
\begin{equation}
\label{equ1} \epsilon(\omega)=\epsilon_{\infty} \prod_j
\frac{\omega_{Lj}^2-\omega^2-i\gamma_{Lj}\omega}{\omega_{Tj}^2-\omega^2-i\gamma_{Tj}\omega}
\end{equation}
Here $\omega_T$, $\omega_L$, $\gamma_T$ and $\gamma_L$ correspond to
longitudinal (L) and transversal (T) optical eigenfrequency and damping of the
$j$th mode, respectively. $\epsilon_{\infty}$ is the dielectric constant
determined by electronic processes alone. In the paramagnetic phase ($T>T_N$)
we have a single mode and $j=1$. For temperatures $T<T_N$, in the AFM phase, we
had to use $j=3$ to describe the data consistently. At normal incidence the
dielectric function is related to the reflectivity via
\begin{equation}
\label{equ2}
R(\omega)=\left|\frac{\sqrt{\epsilon(\omega)}-1}{\sqrt{\epsilon(\omega)}+1}\right|^2.
\end{equation}
In this formalism the dielectric strength of a given mode is determined by
\begin{equation}
\label{equ3}
\Delta\epsilon=\epsilon_{\infty}\frac{\omega_L^2-\omega_T^2}{\omega_T^2}
\end{equation}
and the effective plasma frequency, which is related to the effective charges,
follows from
\begin{equation}
\label{equ4}
\Omega^2=\epsilon_{\infty}(\omega_L^2-\omega_T^2)=\frac{N}{V}\frac{\epsilon_{\infty}}{\epsilon_{vac}}\frac{(Z^*e)^2}{\mu}.
\end{equation}
Here $N/V$ is the number of ion pairs per unit cell and $\epsilon_{vac}$ the
permittivity of free space. $Z^*e$ is the effective charge, which should be
close to~2 in mainly ionic MnO, and $\mu$ is the reduced mass of the ions
involved in the lattice vibration.

\begin{figure}
\includegraphics{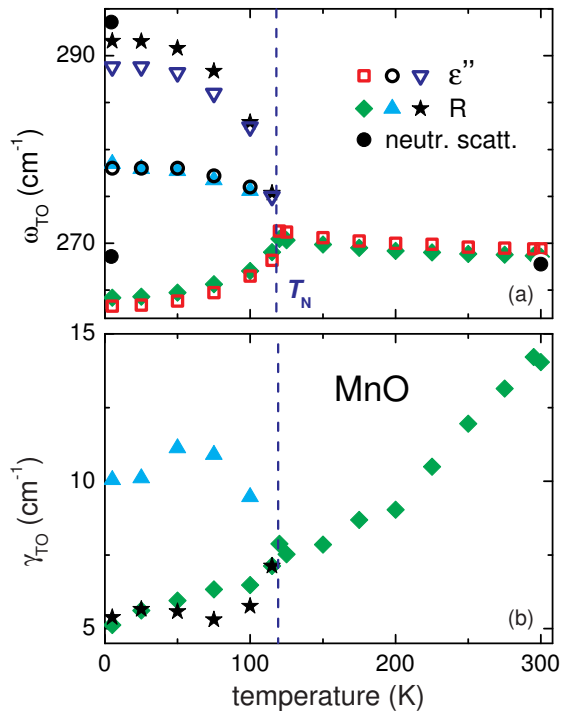}
\caption{\label{fig2}(Color online) Eigenfrequencies (upper frame) and damping
constants (lower frame) of transverse optical modes of MnO. Closed symbols have
been determined from the fits of $R(\omega)$. Empty symbols were read off from
the peak maxima in $\epsilon''$. Full circles indicate neutron scattering
results (Ref.~\onlinecite{chung}).}
\end{figure}

The reflectivity spectra have been fitted using Eqs.~(\ref{equ1})
and~(\ref{equ2}) utilizing a fit routine which has been developed by A.
Kuzmenko.~\cite{kuzmenko} The reflectivity spectrum is fully determined by the
transversal and longitudinal eigenfrequencies and damping constants, which are
treated as free parameters (four-parameter fit). In the AFM state the
dielectric strength and the effective plasma frequency has to be calculated for
each mode $j$ separately (see e.g. Ref.~\onlinecite{rudolf2007}). The
electronic dielectric constant $\epsilon_{\infty}$ has been determined from the
reflectivity measured up to 8000~cm$^{-1}$ ($\sim 1~$eV). $\epsilon_{\infty}$
obtained from these fits was found to scatter between 4.7 and 5.0, with no
systematic temperature variation. Hence for all further analyses we used an
average value of 4.85 for all temperatures from $5~\textrm{K} < T <
305~\textrm{K}$. In addition, because of the strong anharmonic effects we
excluded the reflectivity regime between 350 and 480~cm$^{-1}$ in our fit
routine at all temperatures and found, that this restriction provides a more
reliable fit. The results of the best fits are shown as solid lines in
Fig.~\ref{fig1} and provide a good description of the reststrahlen spectrum of
MnO. Even in the magnetically ordered phase, where a clear splitting of the
modes can be observed, a fit using three eigenmodes yields a convincing
description of the experimentally observed reflectivity (see inset in
Fig.~\ref{fig1}). From these fits we determined all transverse and longitudinal
eigenmodes and damping constants as a function of temperature.

Fig.~\ref{fig2}(a) and (b) show eigenfrequencies (upper frame) and dampings
(lower frame) of the transverse eigenmodes as a function of temperature. In the
paramagnetic phase $\omega_{TO}$ reveals a slight increase on decreasing
temperature as expected for an anharmonic solid. But more importantly, as is
impressively documented in Fig.~\ref{fig2}(a), a clear splitting of the
eigenfrequencies can be observed in the AFM phase. At 5~K we found three
transverse eigenmodes with frequencies close to 264, 278 and 292~cm$^{-1}$.
Hence the overall splitting amounts to almost 30~cm$^{-1}$.

In the paramagnetic phase the temperature dependence of the damping decreases
as expected from anharmonic like phonon-phonon interactions
[Fig.~\ref{fig2}(b)]. Below $T_N$ the damping of the peak with the lowest
frequency continues to decrease, as does the damping of the mode with the
highest frequency and at the lowest temperatures the damping amounts about 2\%
of the mode eigenfrequency. The damping of the peak in between of these two
modes, with a very low dielectric strength, is of the order of 10~cm$^{-1}$.
However, one has to be aware that the experimental uncertainties for this mode
are rather large. We also would like to mention that at room temperature the
longitudinal eigenmode is characterized by an eigenfrequency
$\omega_{LO}=556$~cm$^{-1}$ and a damping constant $\gamma_{LO}=21$~cm$^{-1}$.

\begin{figure}
\includegraphics{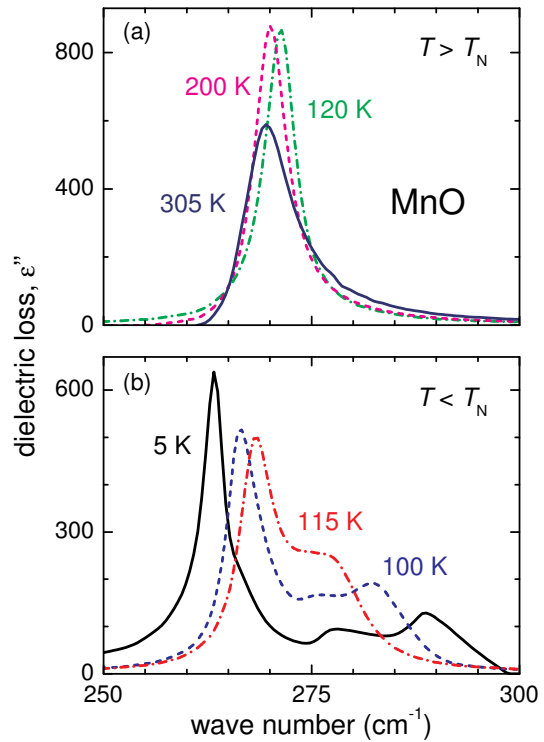}
\caption{\label{fig3}(Color online) Dielectric loss of MnO at various
temperatures calculated via Kramers-Kronig transformation from the
reflectivity. The shift of the main mode and the emergence of new peaks with
low intensity below $T_N$ is clearly documented. The upper frame shows the
dielectric loss in the paramagnetic phase, the lower frame in the AFM state.}
\end{figure}

To arrive at an independent check of the quality and accuracy of the
temperature dependence and splitting of the mode eigenfrequencies as shown in
Fig.~\ref{fig2}(a) we compared these fit results to the maxima of the
dielectric loss function. To do so, we calculated the dielectric loss from the
measured reflectivity spectra (70--8000~cm$^{-1}$), with a constant
extrapolation towards low frequencies and a smooth $\omega^{-h}$ extrapolation
towards high frequencies with exponents $h$ ranging from 0.09 to 0.22. In
Fig.~\ref{fig3} we show the dielectric loss between 250 and 300~cm$^{-1}$
determined by this procedure. For $T > T_N$, in the paramagnetic phase, we
detected a single loss peak indicating one well defined transverse optical
phonon as expected in the cubic NaCl-type phase. Below $T_N$ this mode reveals
significant splitting into two modes at 115~K and finally, at lower
temperatures, even into three modes which, at 5~K, are located at 263, 278 and
289~cm$^{-1}$. These eigenfrequencies, as determined from the peak maxima of
the loss peaks, are plotted in the upper frame of Fig.~\ref{fig2} and we find
excellent agreement between the eigenfrequencies as determined out of
$R(\omega)$ by the four-parameter fit routine and those derived from the loss
peaks.

Our far-infrared results on the splitting of the modes have to be compared with
recent neutron scattering results, which determined one mode at 300~K close to
268~cm$^{-1}$ and two modes at 4.3~K, located at 268.6 cm$^{-1}$ and 293.6
cm$^{-1}$.~\cite{chung} The eigenfrequencies of these modes are indicated in
Fig.~\ref{fig2}(a). The two low-temperature modes obviously correspond to the
two more intense modes as documented in Fig.~\ref{fig3}. The small mode in
between probably is unobservable in neutron scattering due to the weak
intensity and the limited experimental resolution. However, the observation of
a splitting into three modes may elucidate the underlying mechanism driving
this phonon splitting and may put serious constraints on model calculations.

\begin{figure}
\includegraphics{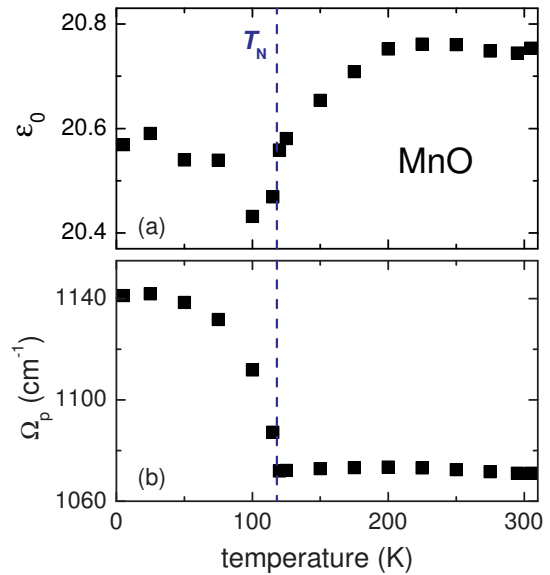}
\caption{\label{fig4}(Color online) Temperature dependence of the static
dielectric susceptibility $\epsilon_0$ (upper frame) calculated via the
Lyddane-Sachs-Teller relation and a constant average $\epsilon_{\infty}$ of
4.85. The lower frame shows a steep increase in the overall plasma frequency
below $T_N$.}
\end{figure}

Using Eqs.~(\ref{equ3}) and~(\ref{equ4}) we calculated the static dielectric
constant $\epsilon_0$ and the ionic plasma frequency, the results are shown in
Fig.~\ref{fig4}. For $T>T_N$, the ionic plasma frequency is almost temperature
independent with an average value of $\Omega_p = 1092$~cm$^{-1}$
[Fig.~\ref{fig4}(b)]. Below the N\'{e}el temperature the effective plasma frequency
exhibits an order parameter like increase and saturates close to 1140~cm$^{-1}$
at low temperatures. This increase of the plasma frequency of more than 6\% in
the magnetically ordered state indicates significant charge transfer processes.
The temperature dependence of the static dielectric constant, as calculated by
Eq.~\ref{equ3}, is shown in the upper panel of Fig.~\ref{fig4}. At low
temperatures we summed up the dielectric strength of all three eigenmodes.
$\epsilon_0$ slightly decreases from 20.75 at room temperature to 20.6 at 5~K,
with a dip-like suppression close to $T_N$. The results of the static
dielectric constant can be compared with measurements of the dielectric
constants at radio frequencies utilizing canonical dielectric
spectroscopy.~\cite{seehra} At 270~K, zero external pressure and depending on
polarization these authors determined dielectric constants between 18.05 and
18.92, which is significantly lower than our value. In addition they report a
slight decrease of the dielectric constant below the magnetic ordering
temperature while we find a cusp like suppression close to $T_N$. These
discrepancies remain unexplained.

Longitudinal and transverse eigenfrequencies, as well as $\epsilon_0$ and
$\epsilon_{\infty}$, are in good agreement with earlier
results.~\cite{plendl,kinney,mochizuki} However, one has to keep in mind that
these earlier results have been obtained by analyzing the reflectivity spectrum
using a model involving two damped oscillators. In these earlier
investigations, the longitudinal eigenfrequency has been calculated from the
Lyddane-Sachs-Teller (LST) relation:
\begin{equation}
\label{equ7}
\frac{\epsilon_0}{\epsilon_{\infty}}=\left(\frac{\omega_L}{\omega_T}\right)^2
\end{equation}
In our analysis the LST relation automatically is fulfilled. Hence, in the
present model a good fit is derived with the use of five parameters [see
Eq.~(\ref{equ1})], while the models used in Refs.~\onlinecite{plendl}
and~\onlinecite{mochizuki} utilize a set of seven parameters. In addition, it
is clear that the second mode is rather unphysical or has to be assigned to a
zero-wave-vector two-phonon process, which should be observable throughout the
Brillouin zone. At the zone boundary only the density of states is maximum,
producing the hump around 470~cm$^{-1}$ as observed in Fig.~\ref{fig1}. Our
result of the transverse optical phonon is also in good agreement with early
neutron scattering results,~\cite{haywood2,haywood4,wagner1} while it seems
that the longitudinal phonon mode has been underestimated by neutron scattering
techniques.~\cite{haywood4} In Ref.~\onlinecite{haywood4} the eigenfrequencies
of the LO branch at the zone center has been determined as 484~cm$^{-1}$
compared to 556~cm$^{-1}$ observed in this work.

An analysis of the ionic plasma frequency also allows clear statements
concerning the ionicity of the bonds in MnO. From detailed
electron-density-distribution studies utilizing gamma-ray
spectroscopy,~\cite{jauch} it has been concluded that the bonding is purely
ionic. We can calculate the ionic plasma frequency assuming an ideal ionic
valence $Z = \pm 2$ for both ions, which results in an effective plasma
frequency of 1873~cm$^{-1}$, which is significantly higher than the
experimentally observed value, $\Omega_p = 1077~$cm$^{-1}$ at room temperature.
This implies that $Z^*/Z \sim 0.58$, which still signals predominantly ionic
bonding, however with distinct covalent contributions. The effective charges
are almost independent of temperature for $T > T_N$ but reveal significant
charge transfer in the AFM phase.

\section{Concluding remarks}

In this work we document a detailed far-infrared study of the phonon properties
of MnO with special emphasis on the spin-phonon coupling and on the phonon
splitting in the AFM phase for $T < 118$~K. MnO reveals a $d^5$ configuration
with a spin-only value of $S = 5/2$, it is an electronically strongly
correlated material and a prototypical Mott-Hubbard insulator. We provide clear
experimental evidence for\\
i) a significant influence of spin ordering on eigenfrequencies, damping and
effective plasma frequencies. Specifically below the antiferromagnetic phase
transition we find an increase of the effective charge by more than 5\%. This
fact reflects strong charge transfer processes and/or strong changes in the
dynamic polarizability.\\
ii) Below $T_N$ we find a splitting of the phonon modes into three modes with
an overall splitting of as much as 30~cm$^{-1}$. The observed splitting
compares well with neutron scattering results by Chung {\it et
al.},~\cite{chung} who, however, observe a splitting into two modes only. In
neutron scattering the ratio of the integrated intensities of the two peaks was
approximately 2:1 in favor of the lower mode. Also the far-infrared results at
5~K [Fig. 3(b)] show an intense low-frequency peak with approximately 2/3 of
the total intensity. It is unclear if the third peak in neutron scattering was
unobservable due to a weak intensity or due to a finite resolution. It is
possible that this low-intensity peak at 280~cm$^{-1}$ reflects the symmetry in
the antiferromagnetic state which should be lower than rhombohedral.

It is unclear if the present experiments allow for a critical review on
spin-Jahn-Teller transitions. MnO is a strongly correlated material with the
spins residing on an fcc lattice. It reveals only weak magnetic frustration $(f
\approx 5)$ and exhibits giant exchange striction at $T_N$. One is apt to think
that symmetry lowering by exchange striction alone is enough to explain the
phonon splitting. However, it has been clearly demonstrated by Massidda {\it et
al.}~\cite{massidda} that the structural distortion at $T_N$ only explains
phonon splittings of the order of less than 1\%. The calculated magnetic order
induced splitting of the optical phonon modes at the zone center was estimated
to range model dependent from 3 to 10\%. Experimentally we observed effects
larger than 10\%. We conclude that MnO has to be described within the framework
of spin-driven Jahn-Teller transitions, developed to describe similar
observations in a variety of spinel compounds.

\begin{acknowledgements}
This research has partly been supported by the Deutsche Forschungsgemeinschaft
through the German Research Collaboration SFB~484 (University of Augsburg).
\end{acknowledgements}

\end{document}